\documentclass[12pt,a4paper]{article}
\usepackage[cp1251]{inputenc}
\usepackage[russian,english]{babel}
\usepackage{graphicx}
\usepackage{amsmath}
\usepackage{indentfirst}
\usepackage{amssymb}

\lccode`\-=`\-
\date{}

\large
\title{Darboux transformation of boundary conditions\\ of regular Dirac Sturm---Liouville problem}
\author{Ekaterina Pozdeeva\footnote{Also Department of Physics, Moscow State
Mining University, 6 Leninsky prospect, Moscow, 119991, Russia;
  E-mail: ekatpozdeeva@mail.ru} and Alexander Tarasov\\Joint Institute for Nuclear Research,\\
  141980, Dubna, Moscow region, Russia}
 
\begin{document}
\selectlanguage{english} \maketitle
\begin{abstract}
It  is shown that boundary conditions of the Darboux transformed
Dirac Sturm---Liouville problem are always zero-valued independently
on boundary conditions of initial  problem.
\end{abstract}

\section{Introduction}
The study of the properties of the Darboux transformation
\cite{D,M,BS1,BS2,R}(or SUSY) attracts more and more attention  in
the past two decades. Recent the developments in this area are
reflected in the special issue of J. Phys. A {\bf 34} where the
references to the early paper can be found.

Despite in depth studies, many problems in given field  have yet to
be solved.  In particular relationship between  Green functions of
two Hamiltonians being SUSY partners have been studied only in few
paper \cite{Su,SSP,SP1,SP2,SP3}. All these paper deal with the Green
functions of Schr\"o\-din\-ger Hamiltonians.

The attempts to generalize  some of these results to the case of
Dirac problem have been only recently \cite{P1,P2,P3,P4}. They lead
to the necessity of consideration of the Darboux transformation of
boundary conditions of the initial  Sturm---Liouville problem.

In the paper \cite{P4} this problem have been considered for some
particular cases,  that admit the analytical solution of the
Sturm---Liouville problem. Considered in \cite{P4} the boundary
values of the Darboux transformed  solutions of these problem are
always equal zero.

The last means that both components of the spinorial solution on the
ends of an interval, on which  leads  to the natural assumption that
this conclusion is also valid in general case, but not only in
particular cases, considered in \cite{P4}.

The proving of this statement is the aim of the present paper. In
Section 2 we derive two different forms of the relation between the
solutions of the initial and Darboux transformed Dirac
Sturm---Liouville problems. In Section 3 we use  these  relation to
prove the main statement of this paper. In Section 4 we analyze the
consequence of obtained results for the  relation between  the Green
functions  of the two SUSY Dirac partners.

\section{Two representations of Darboux transformed solutions of the
Dirac equation}

We consider the solutions of the one-dimensional two-component Dirac
equation
\begin{eqnarray}
  h_0\psi(x)&=&E\psi(x), \qquad  h_0=\gamma\partial_x+V(x),\qquad \gamma=i\sigma_2,\\
  V^+(x)&=&V(x), \qquad \psi(x)=(\psi_1(x),\psi_2(x))\nonumber
\end{eqnarray}
that satisfy the homogenous boundary conditions on the ends of a
close interval $[a,b]$
\begin{eqnarray}
\psi_1(a)\sin(\alpha)+\psi_2(a)\cos(\alpha)&=&0,\\
\psi_1(b)\sin(\beta)+\psi_2(b)\cos(\beta)&=&0.
\end{eqnarray}

It is evidently that  the eigenvalue spectrum of this problem is
discrete.

Let $u(x)$ and $v(x)$ (with the eigenvalues $\lambda_1$, $\lambda_2$
correspondingly)  are the solutions of this boundary problem such
that matrix\begin{eqnarray}
  u(x)&=&\left(%
\begin{array}{cc}
  u_1(x)& v_1(x)\\
  u_2(x) & v_2(x)\\
\end{array}%
\right)
\end{eqnarray}
is nonsingular inside the open interval $(a,b)$.

Consider the Darboux transformation
\begin{eqnarray}
  \label{7} \widetilde{\psi}&=&L\psi,\qquad  L=\partial_x-u_xu^{-1},\qquad u_x=\partial_xu(x).
\end{eqnarray}
If $\psi$ in (\ref{7}) satisfy the equation
\begin{eqnarray}
  h_0\psi&=&E\psi,
\end{eqnarray}
then $\widetilde{\psi}$ in (\ref{7}) satisfy the equation
\begin{eqnarray}
  h_1\widetilde{\psi}&=&E\widetilde{\psi},
\end{eqnarray}
where
\begin{eqnarray}
h_1&=&h_0+[\gamma,u_xu^{-1}].
\end{eqnarray}
Thus, $h_0$ and $h_1$ are SUSY  partners.

The Green function of an initial Sturm---Liouville problem  is the
$2\times 2$ matrix that satisfy an inhomogeneous
equation\begin{eqnarray}
  (h_0-E)G(x,t;E)&=&\delta(x-y)I,
\end{eqnarray}
where  $I$ is the $2\times 2$ unity matrix.

It can be construct  from two solutions (``left'' and ``right'') of
the Dirac equation
\begin{eqnarray}
  h_0\psi_{L(R)}(x,E)&=&E\psi_{L(R)}
\end{eqnarray}
which satisfy following boundary conditions
\begin{eqnarray}
\psi_{L1}(a)\cos(\alpha)+\psi_{L2}(a)\sin(\alpha)&=&0,\\
\psi_{R1}(b)\cos(\beta)+\psi_{R2}(b)\sin(\beta)&=&0.
\end{eqnarray}
Then \begin{eqnarray}
  \label{16} G_{ik}&=&[\psi_{Li}\psi_{Rk}\theta(x-y)+\psi_{Ri}\psi_{Lk}\theta(y-x)]/W\{\psi_L,\psi_R\},
\end{eqnarray}
 $$ W\{\psi_L,\psi_R\}= det\left|%
\begin{array}{cc}
  \psi_{L1}(x) & \psi_{R1}(x) \\
  \psi_{L2}(x) & \psi_{R2}(x) \\
\end{array}%
\right|,$$
\begin{equation}
\frac{dW\{\psi_L,\psi_R\}}{dx}=0.
\end{equation}
The boundary conditions for components of $G_{ik}$ are the
following:
\begin{eqnarray}
G_{1k}(a,y,E)\cos(\alpha)+G_{2k}(a,y,E)\sin(\alpha)&=&0,\\
G_{i1}(x,b,E)\cos(\beta)+G_{i2}(x,b,E)\sin(\beta)&=&0.
\end{eqnarray}

The expression for the Green function of SUSY partner of $h_0$ can
be obtained from (\ref{16}) by the substitution
\begin{eqnarray}
  \psi_{L,R}(x)\rightarrow \widetilde{\psi}_{L,R}(x)=L\psi_{L,R},
\end{eqnarray}
where the operator $L$ is defined by (\ref{7}).

The boundary conditions of the Darboux transformed problem need be
derived from boundary condition of the initial problem and
properties of $L$ the operator $L$.

The equation (\ref{7}) in components reads:
\begin{eqnarray}
 \label{22} \widetilde{\psi}_1&=&\psi'_1-\frac{a}{D}\psi_1-\frac{b}{D}\psi_2,\\
  \label{23}\widetilde{\psi}_2&=&\psi'_2-\frac{d}{D}\psi_1-\frac{c}{D}\psi_2,
\end{eqnarray}
\begin{equation}
  a = v_2u'_1-u_2v'_1,\quad
  b = u_1v'_1-v_1u'_1,
\end{equation}
\begin{equation}
  d = v_2u'_2-u_2v'_2,\quad
  c = u_1v'_2-v_1u'_2,
\end{equation}
\begin{equation}
  D =u_1v_2-u_2v_1.
\end{equation}

For the following the another expressions for the components
$\widetilde{\psi}_{1,2}$, that don't  contain derivations.

If $\psi$ is the arbitrary solution of equation $h_0\psi=E\psi$,
then taking into account that matrix $u$ obey the equation
\begin{eqnarray}
  h_0u&=&u\Lambda, \qquad \Lambda=\left(%
\begin{array}{cc}
  \lambda_1 & 0 \\
  0 & \lambda_2 \\
\end{array}%
\right),
\end{eqnarray}
it is easy to prove that
\begin{eqnarray}
  \label{30} \widetilde{\psi}_1&=&-E\psi_2+\frac{\overline{a}}{D}\psi_1+\frac{\overline{b}}{D}\psi_2,\\
 \label{31} \widetilde{\psi}_2&=&E\psi_1-\frac{\overline{d}}{D}\psi_1-\frac{\overline{c}}{D}\psi_2,
\end{eqnarray}
\begin{equation}
  \overline{a} = (\lambda_1-\lambda_2)u_2v_2,\quad
  \overline{b} = \lambda_2v_2u_1-\lambda_1u_2v_1,
\end{equation}
\begin{equation}
\overline{c} = (\lambda_2-\lambda_1)u_1v_1,\quad
  \overline{d} = \lambda_1u_1v_2-\lambda_2v_1u_2,
\end{equation}
with the help of eqs. (\ref{22}), (\ref{23}), (\ref{30}), (\ref{31})
in the next section we prove that
\begin{eqnarray}
 \label{37} \widetilde{\psi}_{L1}(a,E)&=&\widetilde{\psi}_{L2}(a,E)=0, \\
 \label{38} \widetilde{\psi}_{R1}(b,E)&=&\widetilde{\psi}_{R2}(b,E)=0.
\end{eqnarray}

\section{Boundary conditions of the Sturm---Liouville problem after
Darboux transformation}

Here  we prove only relations (\ref{37}). The proving of (\ref{38})
is similar.

Put $\psi(x,E)=\psi_L(x,E)$ and introduce the following combinations
of quantities $\psi_{L1}$, $\psi_{L2}$, $u_1$, $u_2$, $v_1$, $v_2$:

\begin{equation}
 \label{1q} \phi_L= \cos{(\alpha)}\psi_{L1}+\sin{(\alpha)}\psi_{L2},
\end{equation}
\begin{equation}
  \varphi_L=-\sin{(\alpha)}\psi_{L1}+\cos{(\alpha)}\psi_{L2},
\end{equation}
\begin{eqnarray}
  g_u &=& \cos{(\alpha)}u_1+\sin{(\alpha)}u_2, \quad
  h_u = -\sin{(\alpha)}u_1+\cos{(\alpha)}u_2,\\
  g_v &=& \cos{(\alpha)}v_1+\sin{(\alpha)}v_2, \quad
  h_v = -\sin{(\alpha)}v_1+\cos{(\alpha)}v_2,\\
\widetilde{\phi}_L&=&
\cos{(\alpha)}\widetilde{\psi}_{L1}+\sin{(\alpha)}\widetilde{\psi}_{L2},\quad
  \widetilde{\varphi}_L=-\sin{(\alpha)}\widetilde{\psi}_{L1}+\cos{(\alpha)}\widetilde{\psi}_{L2},
\end{eqnarray}
From here we see that
\begin{equation}
  \phi_L(a)=g_h(a)=g_v(a)=0,
\end{equation}
\begin{equation}
  D=u_1v_2-u_2v_1=g_uh_v-g_vh_u,\quad D(a)=0,
\end{equation}
\begin{eqnarray}
  \psi_{L1}& =&\cos{(\alpha)}\phi_L-\sin{(\alpha)}\varphi_L, \quad
  \psi_{L2} =\sin{(\alpha)}\phi_L+\cos{(\alpha)}\varphi_L,\\
  u_1&=&\cos{(\alpha)}g_u-\sin{(\alpha)}h_u, \quad
  u_2=\sin{(\alpha)}g_u+\cos{(\alpha)}h_u,
\end{eqnarray}
\begin{eqnarray}
  v_1&=&\cos{(\alpha)}g_v-\sin{(\alpha)}h_v, \\
  \label{2q} v_2&=&\sin{(\alpha)}g_v+\cos{(\alpha)}h_v.
\end{eqnarray}

Then combining equations from (\ref{1q}) to (\ref{2q}) it is easy
derive
\begin{eqnarray}
  \label{54} \widetilde{\phi}_L&=&\phi'_L-\frac{g'_u}{D}(\phi_Lh_v-\varphi_Lg_v)-\frac{g'_v}{D}(-\phi_Lh_u+\varphi_Lg_u),\\
  \label{55} \widetilde{\varphi}_L&=&E\phi-\frac{g_u\lambda_1}{D}(\phi h_v-\varphi_Lg_v)-\frac{g_v\lambda_2}{D}(\phi_Lh_u-\varphi_Lg_u)
\end{eqnarray}
Both numerators and  denominators in these expressions become zero
when $x\rightarrow a$.

So, we need use the L'Hospital rules to open the indefinities
arising in eqs. (\ref{54}), (\ref{55}) when $x\rightarrow a.$

The simple calculations leads to the following results:
\begin{eqnarray}
 \label{53} \widetilde{\phi}_L(a)&=&0, \\
 \label{54} \widetilde{\varphi}_L(a)&=&0.
\end{eqnarray}
Combination of eqs.  (\ref{53}), (\ref{54}) leads to the result
\begin{eqnarray}
  \widetilde{\psi}_{L1}(a)=\widetilde{\psi}_{L2}(a)=0,
\end{eqnarray}
cited in previous Section (see equation (\ref{37})).
\section{Conclusion}

In the paper \cite{P4} it has been proved that for arbitrary initial
Dirac Sturm---Liouville problem the following  integral relation is
valid:
\begin{eqnarray}
&&tr\int_a^b(\widetilde{G}(x,x,E)-G(x,x,E))dx=\frac{1}{E-\lambda_1}+\frac{1}{E-\lambda_2}-\Delta,\\
&&\Delta=\frac{\widetilde{\psi}_{L1}(a)\phi_{R1}(a)+\widetilde{\psi}_{L2}(a)\phi_{R2}(a)-
  \widetilde{\phi}_{R1}(b)\psi_{L1}(b)-\widetilde{\phi}_{R2}(b)\psi_{L2}(b)}{W\{\widetilde{\phi}_L,\widetilde{\psi}_R\}}.
\end{eqnarray}
Here $G(x,y,E)$ is the Green function of the initial
Sturm---Liouville problem, $\widetilde{G}(x,y,E)$ is the Green
function of the Darboux transformed problem.

From eqs. (\ref{37}), (\ref{38})  it follows that $\Delta=0$ and the
trace of difference of the Green functions  of the
Hamiltonians-superpartners are equal sum of two pole terms.

\end{document}